\documentclass[aps,prl,showkeys,nofootinbib,superscriptaddress,twocolumn]{revtex4-2}

\usepackage{enumerate}
\usepackage{amsmath,slashed,tensor,amssymb,epsfig,amsthm,bm,graphicx,graphics,slashed}
\usepackage[caption=false]{subfig}
\usepackage{hyperref}
\usepackage[utf8]{inputenc}
\usepackage[top=1.5cm, bottom=1.5cm, left=2cm, right=2cm]{geometry}

\usepackage{makecell}
\usepackage{microtype}
\usepackage[font=footnotesize,labelfont=bf]{caption}

\newcommand{\be}{\begin{equation}}\newcommand{\ee}{\end{equation}}
\newcommand{\bea}{\begin{eqnarray}}\newcommand{\eea}{\end{eqnarray}}
\newcommand{\brr}{\begin{array}}\newcommand{\err}{\end{array}}
\newcommand{\bit}{\begin{itemize}}\newcommand{\eit}{\end{itemize}}
\newcommand{\ben}{\begin{enumerate}}\newcommand{\een}{\end{enumerate}}

\newcommand{\bbm}{\begin{bmatrix}}\newcommand{\ebm}{\end{bmatrix}}
\newcommand{\ba}{\begin{array}}
\newcommand{\ea}{\end{array}}
\newcommand{\G}{\textbf}

\newtheorem{mydef}{Definition}
\newtheorem{Lemma}{Lemma}
\newcommand{\bd}{\begin{mydef}} \newcommand{\ed}{\end{mydef}}
\newcommand{\bthe}{\begin{theorem}} \newcommand{\ethe}{\end{theorem}}
\newcommand{\ble}{\begin{Lemma}} \newcommand{\ele}{\end{Lemma}}

\def\intx{\int \!\!\mathrm{d}^3 {\G x}}

\def\lan{\langle}
\def\lf{\left}

\def\non{\nonumber}\def\pa{\partial}\def\ran{\rangle}

\def\ri{\right}
\def\al{\alpha}

\def\si{\sigma}
\def\om{\omega}

\def\1{{_{1}}}\def\2{{_{2}}}

\def\noHe0{:\;\!\!\;\!\!:H_e(0):\;\!\!\;\!\!:}
\def\noHm0{:\;\!\!\;\!\!:H_\mu(0):\;\!\!\;\!\!:}

\def\lan{\langle}
\def\lf{\left}

\def\non{\nonumber}
\def\pa{\partial}\def\ran{\rangle}

\def\ri{\right}

\def\al{\alpha}

\def\si{\sigma}
\def\om{\omega}

\def\1{{_{1}}}\def\2{{_{2}}}

\begin{document}

\title{Leggett--Garg inequalities 
in the quantum field theory of neutrino oscillations}
\author{Massimo~Blasone}
\email{blasone@sa.infn.it}
\affiliation{Dipartimento di Fisica, Universit\`a di Salerno, Via Giovanni Paolo II, 132 84084 Fisciano, Italy \&
 INFN Sezione di Napoli, Gruppo collegato di Salerno, Italy}
\author{Fabrizio~Illuminati}
\email{filluminati@unisa.it}

\affiliation{Dipartimento  di  Ingegneria Industriale, Universit\`a di Salerno, Via  Giovanni  Paolo  II, 132  I-84084  Fisciano  (SA),  Italy \& INFN,  Sezione  di  Napoli,  Gruppo  collegato  di  Salerno,  Italy.}

\author{Luciano~Petruzziello}
\email{lupetruzziello@unisa.it}

\affiliation{Dipartimento  di  Ingegneria Industriale, Universit\`a di Salerno, Via  Giovanni  Paolo  II, 132  I-84084  Fisciano  (SA),  Italy \& INFN,  Sezione  di  Napoli,  Gruppo  collegato  di  Salerno,  Italy.}

\author{Luca~Smaldone}
\email{smaldone@ipnp.mff.cuni.cz}

\affiliation{Institute of Particle and Nuclear Physics, Faculty  of  Mathematics  and  Physics, Charles  University, V  Hole\v{s}ovi\v{c}k\'{a}ch  2, 18000  Praha  8,  Czech  Republic.}
\date{December 6, 2021}
\vspace{3mm}

\begin{abstract}
We investigate Leggett-Garg temporal inequalities in flavor--mixing processes. We derive an exact flavor--mass uncertainty product and we establish that it is an upper bound to the violation of the inequalities. This finding relates temporal nonclassicality to quantum uncertainty and provides a time analog of the Tsirelson upper bound to the violation of the spatial Bell inequalities. By studying the problem both in the exact field-theoretical setting and in the limiting quantum mechanical approximation, we show that Leggett-Garg inequalities are violated more strongly in quantum field theory than in quantum mechanics.
\end{abstract}
\pacs{}
\keywords{}

\maketitle

\emph{Introduction} -- Neutrino physics covers a very broad spectrum of applications in different fields of scientific investigation~\cite{Bilenky:1978nj, Bilenky:1987ty,Beuthe:2001rc,giunti2007fundamentals}. In particular, the mixed structure of elementary particles has significant implications in the framework of quantum information and quantum resource theory. Indeed, flavor mixing is closely associated with the entanglement of single-particle states and other nonlocal features, both bipartite and multipartite~\cite{ill1,ill2,ill3,ill4,Dixit:2018gjc,PhysRevD.100.055021,ill6}. 



The quantum nature of neutrino oscillations has been probed in the MINOS experiment by observing the violation of a reduced Leggett--Garg type inequality derived in the single-particle quantum mechanical approximation of quantum field theory~\cite{Formaggio2016}. Leggett--Garg inequalities are a central object in fundamental quantum physics and quantum information science, as they can be viewed as the temporal analog of the Bell inequalities quantifying spatial quantum nonlocality~\cite{PhysRevLett.54.857,Brukner2007,Brukner2008,PhysRep2014,Kumari2017}.

Among other possible applications, Leggett--Garg type inequalities have been proposed as a tool to discriminate between Dirac and Majorana fermions by studying neutrino oscillations in matter within phenomenological models of dissipative environments~\cite{Richter:2017toa}. Moreover, different forms of the inequalities have been discussed in the study of three-flavor  neutrino oscillations~\cite{gango,Naikoo:2019gme} and in the investigation of possible non-standard neutrino interactions beyond the Standard Model~\cite{Dixit2021}. 

Surprisingly, despite the fundamental insight on temporal nonclassicality that they could provide at the microscopic scale, Leggett--Garg inequalities in elementary particle physics have been mostly studied in the quantum mechanical approximation rather than in the exact framework of quantum field theory. In this letter, we address for the first time Leggett--Garg inequalities in the full generality of the quantum field theory of neutrino mixing and oscillations \cite{annals,Hannabuss:2000hy,PhysRevD.64.013011,PhysRevD.65.096015,Mavromatos:2012us,Lee:2017cqf}, an ideal playground for the analysis of temporal nonclassicality.

The first result of the present work is that Leggett--Garg temporal inequalities in neutrino physics are intimately related to quantum uncertainty. We obtain this result by observing that in flavor oscillation processes there exists a conserved charge besides the one associated to total lepton number conservation, and that this quantity is a mass--charge operator acting on the flavor vacuum. From the non-commutativity of lepton and mass charges and using the Robertson-Schr\"odinger prescription, we obtain a flavor--mass uncertainty relation that in turn guides us to derive the Leggett--Garg inequality in Wigner form~\cite{Saha2015,Naikoo:2019gme}. 

The second finding, a direct consequence of the first one, is that the flavor-mass uncertainty product is an upper bound to the violation of the Leggett--Garg inequality in neutrino oscillations; this is the first known upper limit to the violation of the temporal inequalities, providing a time analog of the Tsirelson upper bound to the violation of the spatial Bell inequalities. Finally, by comparing the different expressions, we show that the Leggett--Garg inequalities are violated more strongly in quantum field theory than in quantum mechanics.

\emph{Flavor charges and mass-charge operator} -- Consider the Lagrangian density
\be \label{mixlag}
\mathcal{L}(x)\ = \ \begin{pmatrix} \bar{\nu}_e(x) & \bar{\nu}_\mu(x) \end{pmatrix} \, \lf(i\slashed{\pa} -\begin{pmatrix} m_e & m_{e\mu}\\ m_\mu & m_{e\mu} \end{pmatrix}\ri) \, \begin{pmatrix} \nu_e(x) \\ \nu_\mu(x) \end{pmatrix}  \, ,
\ee
which is the free part of the weak interaction Lagrangian in the \emph{flavor} basis~\cite{Blasone2019}. This expression can be diagonalized in the \emph{mass} basis \cite{Bilenky:1978nj}
\bea
\left(
  \begin{array}{c}
    \nu_e (x) \\
    \nu_{\mu}(x) \\
  \end{array}
\right) \ = \ \left(
                \begin{array}{cc}
                  \cos\theta & \sin\theta \\
                  -\sin\theta & \cos\theta \\
                \end{array}
              \right)\left(
                       \begin{array}{c}
                        \nu_1 (x)\\
                         \nu_2 (x) \\
                       \end{array}
                     \right) \, , \label{PontecorvoMix1}
\eea
where $\tan 2 \theta = 2 m_{e\mu}/(m_\mu-m_e)$. The mass and flavor representations are unitarily inequivalent representations of the fermionic anticommutation relations \cite{annals,Hannabuss:2000hy,PhysRevD.64.013011,PhysRevD.65.096015,Mavromatos:2012us,Lee:2017cqf}, and in the following we will work in the flavor basis.

One can verify that the Lagrangian \eqref{mixlag} is invariant under the action of the group
$\mathrm{U}(1)\times\mathrm{U}(1)$~\cite{bigs1}. Hence,
the ensuing conserved charges are
\bea
Q & = & \sum_{\si=e,\mu} \, \intx :\nu^\dag_\si \, \nu_\si: \, , \\[2mm]
Q_M & = & \intx \,  :\begin{pmatrix} \nu^\dag_e & \nu^\dag_\mu \end{pmatrix} \,\begin{pmatrix} m_e & m_{e\mu}\\ m_\mu & m_{e\mu} \end{pmatrix} \, \begin{pmatrix} \nu_e \\ \nu_\mu \end{pmatrix} : \, . 
\eea
Here, $Q$ is the generator of the global phase transformation $\nu_\si \to e^{i \al} \nu_\si$ associated with the \emph{total lepton number} (indeed, it commutes with the neutrino production/detection vertex \cite{Bilenky:1987ty,Blasone2019}), while $Q_M$ is a \emph{mass-charge operator}. The lepton charge can be decomposed as
\bea
Q \ = \ Q_e(t) + Q_\mu(t) \, , 
\eea
where
\bea \label{fcs}
Q_\si(t) & = & \intx \, :\nu^\dag_\si \, \nu_\si: \, , \quad \si \ = \ e,\mu \, ,
\eea
are the flavor lepton charges \cite{Bilenky:1987ty,BHV99}. Therefore, neutrino flavor states $|\nu^r_{\G k,\si}\ran$ with momentum $\textbf{k}$ are defined as eigenstates of such lepton charges at some fixed time.

\emph{Flavor-mass uncertainty relations} -- Based on the above, we can introduce an uncertainty relation stemming from the non-commutativity of the flavor charges $Q_\si$ and $Q_M$. In the Robertson--Schr\"{o}dinger form:
\be \label{fm}
\si^2_Q \, \si^2_M \ \geq \frac{1}{4} \, \lf|\lan \lf[Q_\si(t)\, , \, Q_M\ri]\ran_\si\ri|^2 \, . 
\ee
The above inequality defines a \emph{flavor-mass uncertainty relation}. 
Computing the variances, one finds:
\bea \non
\sigma^2_Q & = & \lan Q^2_{\si}(t)\ran_\si \ - \ \lan Q_{\si}(t)\ran_\si^2 \\[1mm] \label{varq}
& = & \ \mathcal{Q}_{\si\rightarrow \si}(t)\lf(1-\mathcal{Q}_{\si\rightarrow \si}(t)\ri) \, ,\\[2mm]
\sigma^2_M & = & \lan Q^2_M\ran_\si \ - \ \lan Q_M\ran_\si^2 \ = \ m^2_{e\mu} \, .
\eea
The field-theoretical flavor oscillation probability is the expectation value of the flavor charges with respect to a reference flavor state: $\mathcal{Q}_{\si\rightarrow \rho}(t) =  \lan Q_{\nu_\rho}(t) \ran_\si$ \cite{BHV99},
where $\langle \cdots\rangle_\si = \lan \nu^r_{\G k,\si}| \cdots |\nu^r_{\G k,\si}\ran$. 
Explicitly: 
\bea  \non
\mathcal{Q}_{\si\rightarrow \rho}(t)  & = &  \sin^2 (2 \theta)\Big[|U_\G k|^2\sin^2\lf(\om_{\G k}^{_-}t\ri)
 +   |V_\G k|^2\sin^2\lf(\om_{\G k}^{_+}t\ri)\Big]  ,  \\[2mm] \label{oscfor}
\mathcal{Q}_{\si\rightarrow \si}(t)  & = & 1 \ - \ \mathcal{Q}_{\si\rightarrow \rho}(t) \, , \quad \si \neq \rho \, ,
\eea
where $\om_{\G k}^{_{\pm}}\equiv (\om_{\G k,2}\pm\om_{\G k,1})/2$ are combinations of mass-definite energies and
\bea
\label{uk}
|U_\G k| & = & \mathcal{A}_\G k  \lf(1+\frac{|\G k|^2}{\lf(\om_{\G k,1}+m_1\ri)\lf(\om_{\G k,2}+m_2\ri)}\ri)\, ,\\[2mm]
|V_\G k| & = & \mathcal{A}_\G k \lf(\frac{|\G k|}{\om_{\G k,1}+m_1}-\frac{|\G k|}{\om_{\G k,2}+m_2}\ri)\, ,  \\[2mm]
\mathcal{A}_\G k & = & \sqrt{\lf(\frac{\om_{\G k,1}+m_1}{2 \om_{\G k,1}}\ri) \, \lf(\frac{\om_{\G k,2}+m_2}{2 \om_{\G k,2}}\ri)} \, . \label{vk}
\eea
The evaluation of the r.h.s. of Eq.~\eqref{fm} yields
%
%
%
\bea
&& \lf|\lan \lf[Q_\si(t)\, , \, Q_M\ri]\ran_\si\ri| = m_{e \mu} \, \mathcal{C}(t)\\[2mm] \non 
&& = \  m_{e \mu} \sin (2 \theta)\lf|\Big[|U_\G k|^2\sin\lf(2\om_{\G k}^{_-}t\ri)+  |V_\G k|^2\sin\lf(2\om_{\G k}^{_+}t\ri)\Big]\ri| ,
\eea
and thus the quantum field-theoretical flavor--mass uncertainty relation 
\be\label{nonf}
{\mathcal{F}}_{QFT} \ \equiv \ \mathcal{Q}_{\si\rightarrow \si}(t)\lf(1-\mathcal{Q}_{\si\rightarrow \si}(t)\ri) -\frac{1}{4}\mathcal{C}^2(t) \ \geq \ 0 \, .
\ee
In the ultra-relativistic limit $m_i/|\G k| \rightarrow \infty$ (i.e. when $|U_\G k|\rightarrow 1$ and $|V_\G k|\rightarrow 0$), the exact field-theoretical expression $\mathcal{F}_{QFT}$ reduces to
the quantum mechanical approximation $\mathcal{F}_{QM}$:
\be\label{resem}
\mathcal{F}_{QM} \ \equiv \ \mathcal{P}_{\si\rightarrow \si}(t)\lf(1-\mathcal{P}_{\si\rightarrow \si}(t)\ri) -\frac{1}{4}\mathcal{P}^2_{\si\rightarrow \rho}(2t) \ \geq \ 0 \, ,
\ee
where the quantum mechanical oscillation probability is
\be \label{bpfor}
\mathcal{P}_{\si \rightarrow \rho} (t) \ = \ \sin^2 (2 \theta) \, \, \sin^2\lf(\om_{\G k}^{_-}t\ri) \, .
\ee
Therefore, in this regime, Eq.~\eqref{fm} becomes
\be \label{fmunsl}
\mathcal{P}_{\si \rightarrow \rho} (t) \, (1-\mathcal{P}_{\si \rightarrow \rho} (t)) \, \ \geq \ \frac{1}{4}\mathcal{P}_{\si \rightarrow \rho} (2 t) \, , \quad \si \neq \rho \, .
\ee
%
The flavor--mass uncertainty relation yields that the lower bound on the  transition probability product  at a generic time $t$ is set by the transition probability at time $2 t$. 

In the following, we will show that this fact unambiguously establishes the existence of time correlations for flavor neutrinos in terms of Leggett--Garg temporal inequalities and their relation with the flavor--mass uncertainty principle derived above. 

\emph{Leggett--Garg inequalities in quantum field theory} -- Consider an observer performing measurements on a dichotomic variable having outputs $\pm 1$ at different times $t_0,t_1,t_2$. Let $O(t)$ be an operator which quantifies such observable ($O|\pm\ran = m|\pm\ran$, with $m=\pm 1$). One can then introduce temporal relations, analog of the spatial Bell inequalities, that are known as the \emph{Leggett--Garg} temporal inequalities \cite{PhysRevLett.54.857}. Various forms of the latter can then be introduced, by working with the joint probabilities $P(m_i,m_j)$ on the measurements performed at three different times $t_0,t_1,t_2$ \cite{Kumari2017,Saha2015}. 
%

In what follows, we investigate temporal inequalities in the field theory framework for neutrino oscillations. We will establish that temporal inequalities are violated more strongly in quantum field theory than in the quantum mechanical approximation, thus showing that quantum field theory is more nonclassical than quantum mechanics in the time regime. Moreover, we will show that the flavor-mass uncertainty product derived above provides an upper bound to the violation of the Leggett-Garg inequalities, thus realizing the first temporal analog of the Tsirelson upper limit on the violation of the spatial Bell inequalities \cite{Tsirelson1,Tsirelson2,Tsirelson3,Tsirelson4}.

Consider the quantities $O(t)=Q_3(t) \equiv (Q_e(t)-Q_{\mu}(t))$ (i.e. the flavor charges defined in Eq.~\eqref{fcs}). One can verify that $Q_3(t)|\nu^r_{\G k,e}(t)\ran=|\nu^r_{\G k,e}(t)\ran$ and $Q_3(t)|\nu^r_{\G k,\mu}(t)\ran=-|\nu^r_{\G k,\mu}(t)\ran$. Hence, this operator is a dichotomic variable quantifying the neutrino flavor. 
Without loss of generality, let us assume that a muonic neutrino is produced at time $t_0=0$. After that,  two measurements are performed at $t_1=t$ and $t_2=2 t$.

\begin{widetext}

\begin{figure}
\includegraphics[width=14cm]{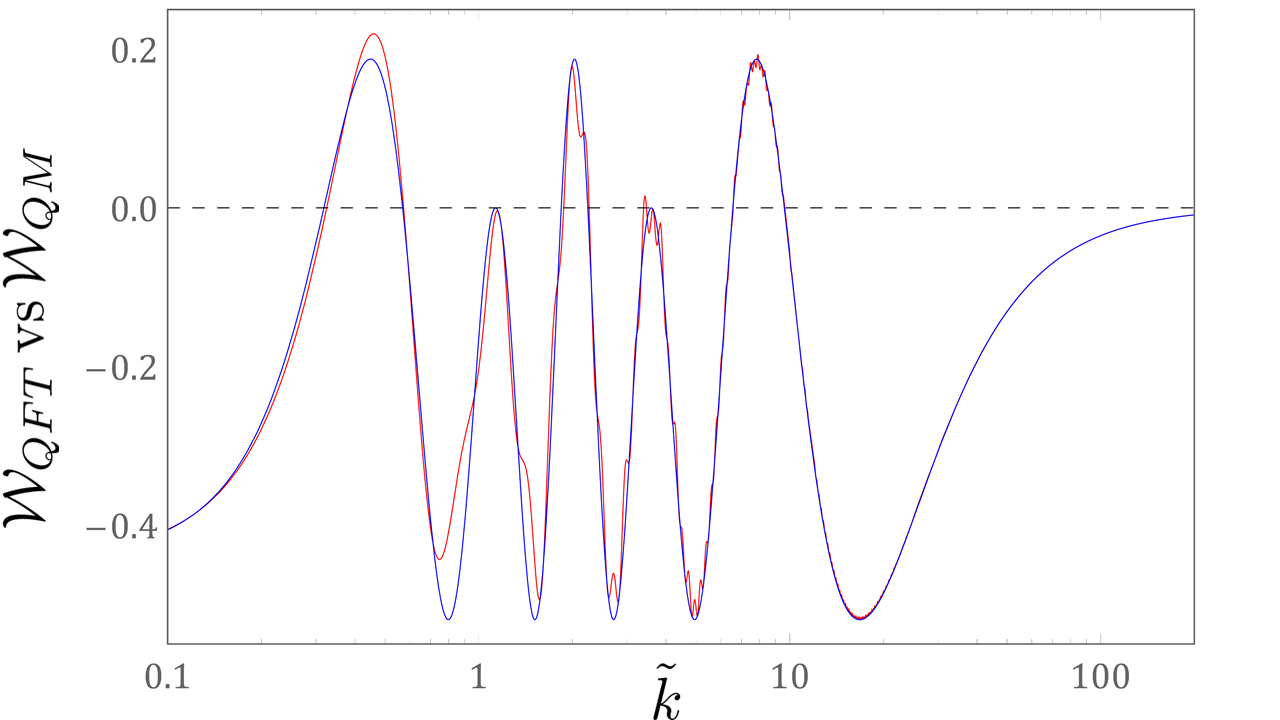}
\caption{Violation of the Leggett--Garg inequality in Wigner form, as a function of $\tilde{k} \equiv |\G k|/\sqrt{m_1 \, m_2}$, in quantum field theory (red line) and in quantum mechanics (blue line), for sample values $m_1=3$, $m_2=20$, $t=1$ and $\theta=\pi/3$. The maximal violation of the inequality is larger in quantum field theory than in quantum mechanics.}
\label{fig}
\end{figure}

\end{widetext}

Then, the general Leggett--Garg temporal inequality in Wigner form \cite{Naikoo:2019gme} reads
%
%
%
\be
P(m_1,m_2)-P(m_0,m_1)-P(-m_0,m_2) \ \leq \ 0 \, ,
\ee
where $m_0=m_1=m_2=1$. After some algebra, in the exact quantum field-theoretical formulation we obtain
\be \label{lgwn}
\mathcal{W}_{QFT} \equiv \mathcal{Q}_{e \rightarrow e} (t) \, \mathcal{Q}_{\mu \rightarrow e} (t)-\mathcal{Q}_{\mu \rightarrow e} (2t) \ \leq \ 0 \, .
\ee
In the ultra-relativistic regime, the limiting quantum mechanical approximation to Eq.~\eqref{lgwn} reads
\be \label{lgwn1}
\mathcal{W}_{QM} \equiv \mathcal{P}_{e \rightarrow e} (t) \, \mathcal{P}_{\mu \rightarrow e} (t)-\mathcal{P}_{\mu \rightarrow e} (2t) \ \leq \ 0 \, ,
\ee
which coincides with the standard form used for the investigation of temporal correlations in neutrino physics~\cite{Naikoo:2019gme}. In Fig.~\ref{fig}, we report $\mathcal{W}_{QFT}$ and $\mathcal{W}_{QM}$ as functions of $|\G k|$, for given values of the masses and of the time $t$ (which here plays the r\^ole of the baseline). The positive absolute maximum of $\mathcal{W}_{QFT}$ is larger than the positive absolute maximum of $\mathcal{W}_{QM}$, showing that the violation of the Leggett--Garg inequality is stronger in quantum field theory than in quantum mechanics. As expected, $\mathcal{W}_{QFT}$ and $\mathcal{W}_{QM}$ coincide in the ultraviolet and in the infrared limits, with a significant deviation occurring in the intermediate regime $|\G k| \approx \sqrt{m_1 \, m_2}$.

\begin{figure}%
\vspace*{-5mm}
    \hspace*{-6mm}
    \subfloat[]{{\includegraphics[width=8.5cm]{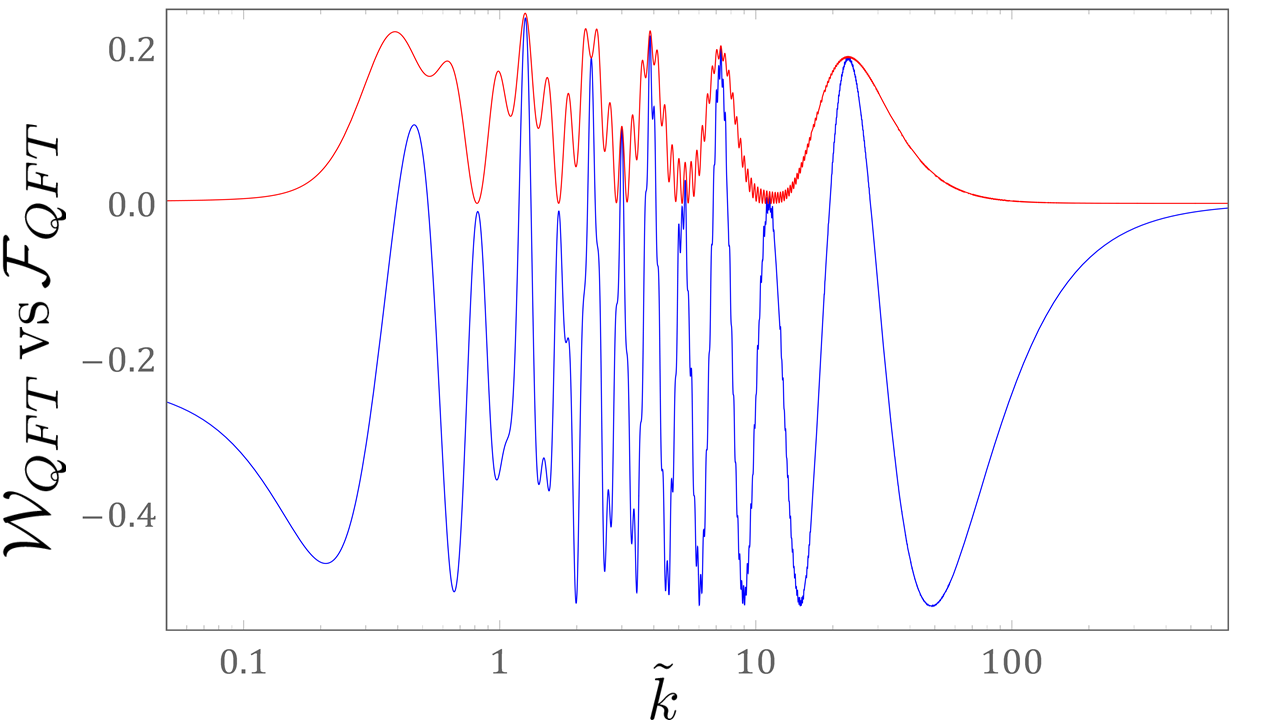} }}%
    \qquad \hspace*{-6mm}
    \subfloat[]{{\includegraphics[width=8.5cm]{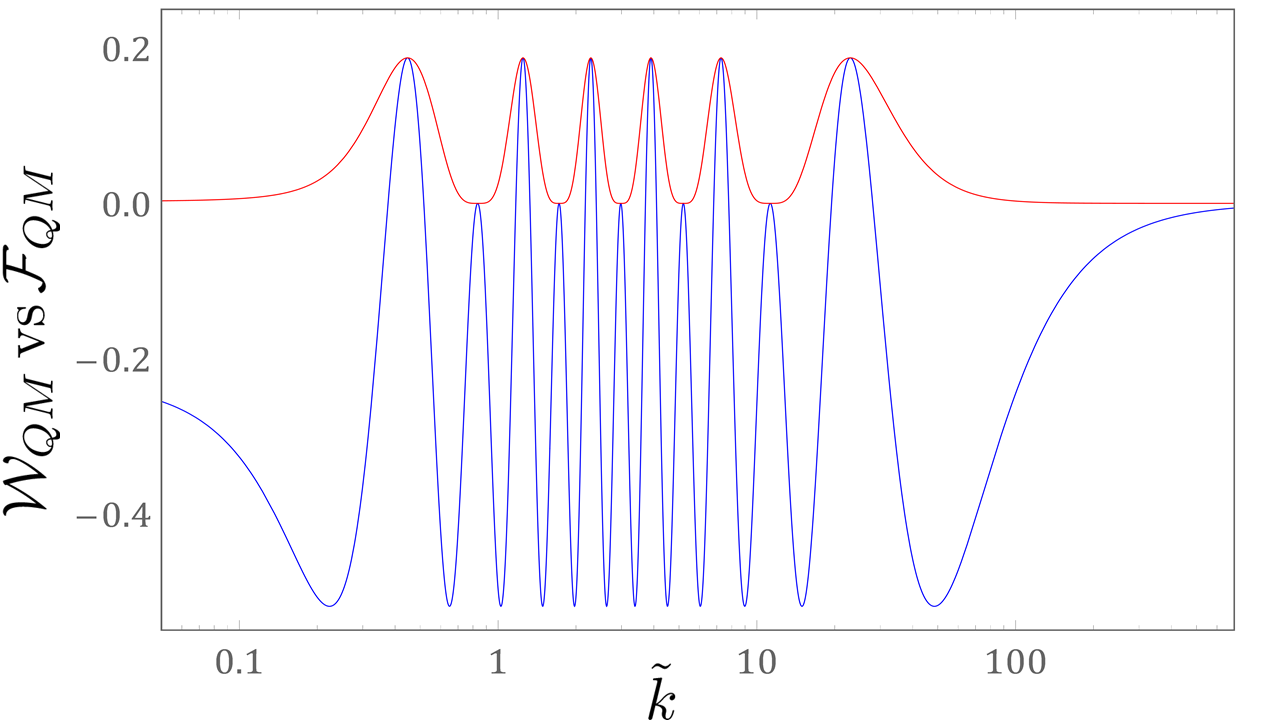} }}%
    \caption{Panel $\mathrm{\textbf{(a)}}$: Behavior of $\mathcal{W}_{QFT}$ (blue line) and of $\mathcal{F}_{QFT}$ (red line) as functions of $\tilde{k} \equiv |\G k|/\sqrt{m_1 \, m_2}$, for sample values $m_1=3$, $m_2=40$, $t=1$ and $\theta=\pi/3$. Panel $\mathrm{\textbf{(b)}}$: Behavior of $\mathcal{W}_{QM}$ (blue line) and of $\mathcal{F}_{QM}$ (red line) as functions of $\tilde{k} \equiv |\G k|/\sqrt{m_1 \, m_2}$, for the same sample values.}%
    \label{fig2}%
\end{figure}

Next, we investigate the relation between the temporal inequalities and the flavor-mass uncertainty~\eqref{fmunsl}. By using Eq.~\eqref{bpfor}, it is immediate to observe that the inequality in Eq.~\eqref{fmunsl} is saturated for $\theta=0$, $\theta=\pi/2$ (no mixing) and $\theta=\pi/4$ (maximal mixing). 
In Fig.~\ref{fig2} we report the behavior of $\mathcal{W}_{QFT}$, $\mathcal{F}_{QFT}$, $\mathcal{W}_{QM}$, and $\mathcal{F}_{QM}$. In all cases, the uncertainty product is an upper bound to the violation of the Leggett--Garg inequality, both in quantum field theory and in quantum mechanics. In the quantum mechanical regime the bound is a trivial consequence of the fact that
\be \label{urd}
\mathcal{F}_{QM} - \mathcal{W}_{QM} \ = \ \frac{3}{4} \, \mathcal{P}_{\si \to \rho}(2 t) \ \geq \ 0 \, .
\ee

\begin{figure}%
    \vspace*{2.5mm}\hspace*{-10mm}
    \subfloat[]{{\includegraphics[width=8.5cm]{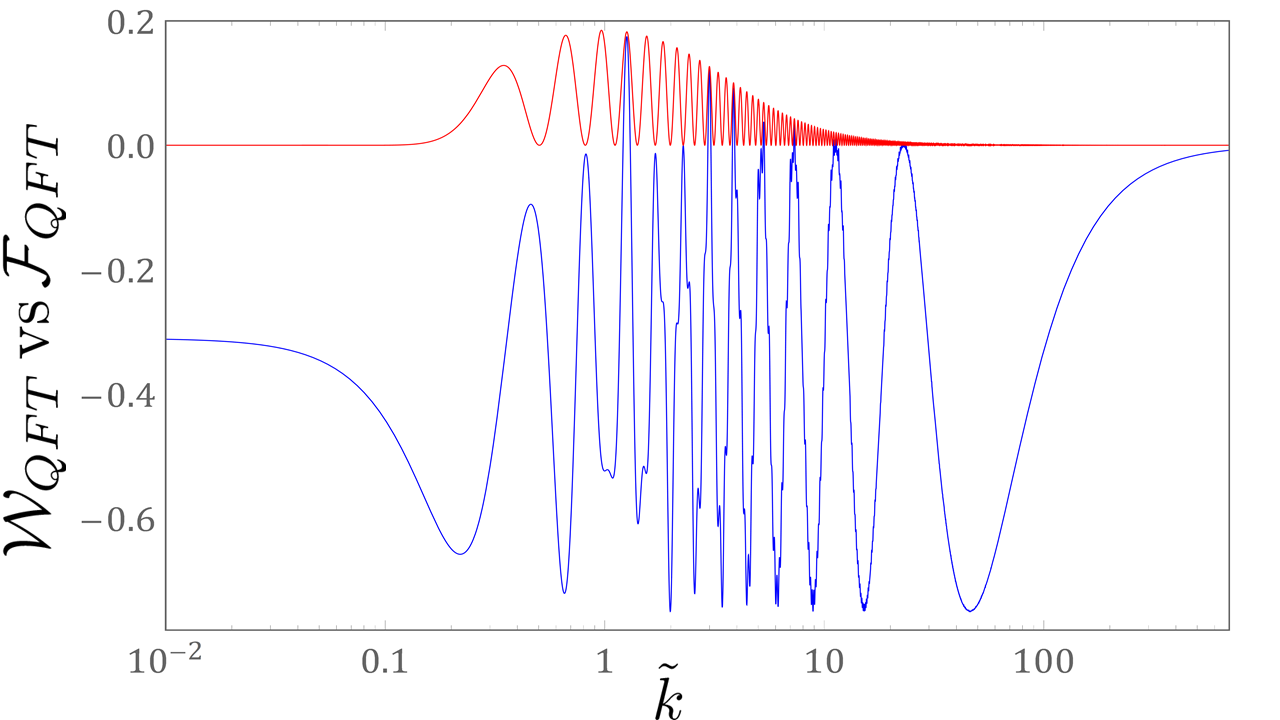} }}%
    \qquad \hspace*{-10mm}
    \subfloat[]{{\includegraphics[width=8.5cm]{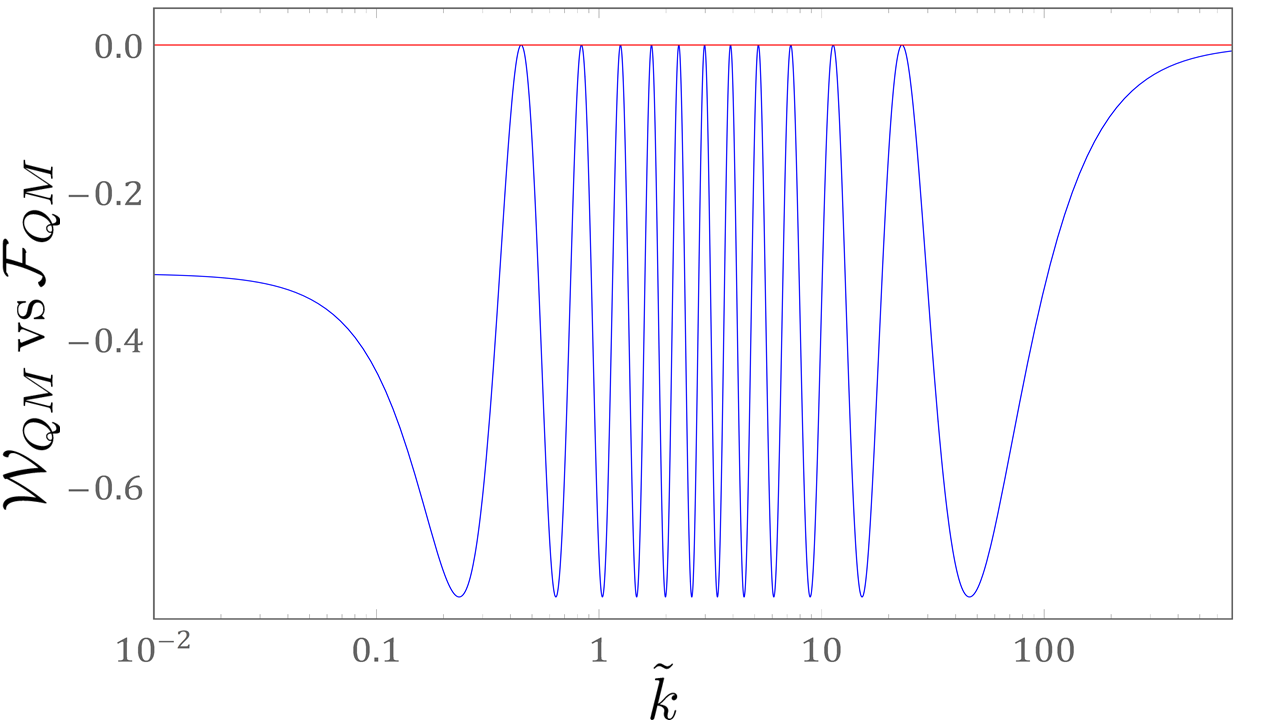} }}%
    \caption{Panel $\mathrm{\textbf{(a)}}$: Behavior of $\mathcal{W}_{QFT}$ (blue line) and of $\mathcal{F}_{QFT}$ (red line) as functions of $\tilde{k} \equiv |\G k|/\sqrt{m_1 \, m_2}$, for sample values $m_1=3$, $m_2=40$, $t=1$ and $\theta=\pi/4$. Panel $\mathrm{\textbf{(b)}}$: Behavior of $\mathcal{W}_{QM}$ (blue line) and of $\mathcal{F}_{QM}$ (red line) as functions of $\tilde{k} \equiv |\G k|/\sqrt{m_1 \, m_2}$, for the same sample values.}%
    \label{fig3}%
\end{figure}

\noindent
For $\theta=\pi/4$, $\mathcal{F}_{QM}$ is identically vanishing and the temporal inequalities are never violated, since $\mathcal{W}_{QM} \leq 0$. Indeed, for this value of the mixing angle neutrino states are near-\emph{classical} coherent states that minimize the flavor-mass uncertainty product.

Turning to quantum field theory, consider the quantity
\bea
4 \lf(\mathcal{F}_{QFT} - \mathcal{W}_{QFT} \ri) & = &  \sin^2(2 \theta) \lf[\sin^2\lf(\om_{\G k}^{_-}t\ri) \, |U_\G k|^2 \lf(4-|U_\G k|^2\ri) \non \ri. \\[2mm]
&+& \sin^2\lf(\om_{\G k}^{_+}t\ri) \, |V_\G k|^2 \lf(4-|V_\G k|^2\ri) \non \\[2mm]
&- & \lf. 2 |V_\G k|^2 \, |U_\G k|^2 \,  \sin\lf(\om_{\G k}^{_+}t\ri) \, \sin\lf(\om_{\G k}^{_-}t\ri) \ri]  .
\eea
Recalling that $|U_\G k|^2+|V_\G k|^2=1$ for any $\textbf{k}$, the above can be rewritten as
\bea
&&  \sin^2(2 \theta)   \lf[\, |V_\G k|^2 \, |U_\G k|^2\lf(\sin\lf(\om_{\G k}^{_-}t\ri)-\sin\lf(\om_{\G k}^{_+}t\ri)\ri)^2 \, \non \ri. \\[2mm]
&&+ \lf. 3 |V_\G k|^2 \, \sin^2\lf(\om_{\G k}^{_+}t\ri) + 3 \, |U_\G k|^2 \, \sin^2\lf(\om_{\G k}^{_-}t\ri) \ri] \geq 0 \, .
\eea
Therefore, $\mathcal{F}_{QFT}$ is always an upper bound to $\mathcal{W}_{QFT}$. Strong violations of the Leggett--Garg inequality in Wigner form can thus occur in quantum field theory even when, as in the case of maximal mixing $\theta=\pi/4$, there is no violation in the quantum mechanical limit, thus definitively certifying the higher degree of temporal nonclassicality of quantum field theory with respect to quantum mechanics, as reported in Fig.~\ref{fig3}. 

Such stronger violation of the Leggett--Garg inequalities in quantum field theory establishes the first temporal analog to the stronger violation of the spatial Bell inequalities, which was proven by Summers and Werner in the full generality of algebraic quantum field theory \cite{Summers:1987ze,Summers:1987fn,Summers:1987fepr}. Those seminal works showed that Bell inequalities are always maximally violated in quantum field theory, even for the vacuum state, thus showing that quantum fluctuations of the vacuum cannot be reproduced with local hidden variable theories. 

In neutrino oscillations, the vacuum of flavor representation (\emph{flavor vacuum}) $|0\ran_{e \mu}$ \cite{annals} is a non-trivial entangled state of mass--neutrinos,
and this feature provides the source of both the pronounced nonlocal behavior of flavor mixing in the temporal regime and the discrepancy between $\mathcal{W}_{QFT}$ and $\mathcal{W}_{QM}$ for $|\textbf{k}|\approx\sqrt{m_1m_2}$. 

\emph{Conclusions and outlook} -- We have investigated the Leggett--Garg temporal inequalities in the quantum field theory of neutrino oscillations. In analogy and in agreement with the general result holding for the spatial Bell inequalities, we have found that the temporal inequalities exhibit a violation that is stronger in quantum field theory than in quantum mechanics.

We have derived a flavor--mass uncertainty relation out of the mass--charge operator and the lepton charges, which holds true beyond the usual non-relativistic regime. Motivated by its peculiar time dependence, we have proved that the flavor--mass uncertainty relation provides an upper bound to the violation of the Leggett--Garg inequalities at all energy scales, establishing a temporal analogue of the Tsirelson's bound for Bell inequalities.

It would be interesting to consider extensions to arbitrary families and generations of elementary particles, in order to probe signatures of new physics beyond the standard model phenomenology and to assess the possibility of applying the present formalism in different settings. For instance, we may think of the axion-photon mixing~\cite{axion1,axion2,axion3,axion4,axion5} and apply the same considerations developed in the present work to quantum optical frameworks. 

In a broader picture, investigating Leggett--Garg inequalities in quantum field theory may have a significant experimental impact, as probing the time nonclassicality of neutrino oscillations can help to shed light on the peculiar nature of such elementary mixed particles. Moreover, it can stimulate novel approaches to the generalization of quantum information science in the relativistic domain and pave the way towards a unified research program on quantum inequalities thanks to the relativistic symmetry of space and time.

\emph{Acknowledgments} -- L. S. acknowledges support from Charles University Research Center (UNCE/SCI/013). F. I. and L. P. acknowledge support by MUR (Ministero dell’Università e della Ricerca) via the project PRIN 2017 ``Taming complexity via QUantum Strategies: a Hybrid Integrated Photonic approach'' (QUSHIP) Id. 2017SRNBRK.

\bibliography{LibraryNeutrino}

\bibliographystyle{apsrev4-2}

\end{document}